\documentstyle[pra,aps,epsfig]{revtex}

\newcommand{\be}{\begin{equation}}
\newcommand{\ee}{\end{equation}}
\newcommand{\bea}{\begin{eqnarray}}
\newcommand{\eea}{\end{eqnarray}}
\newcommand{\bean}{\begin{eqnarray*}}
\newcommand{\eean}{\end{eqnarray*}}
\font\upright=cmu10 scaled\magstep1

\newcommand{\X}{\hbox{\rlap {$X$}\kern 2.2pt {$I$}}}
\newcommand{\identity}{{\upright\rlap{1}\kern 2.0pt 1}}
\newcommand{\half}{\frac{1}{2}}
\newcommand{\bm}{\mathbf}
\newcommand{\lambdabf}{\mbox{\boldmath $\lambda$}}
\newcommand{\mubf}{\mbox{\boldmath $\mu$}}

\newcommand{\sigmabf}{\mbox{\boldmath $\sigma$}}
\begin{document}
\pagestyle{plain}
\preprint{DAMTP-97-4, hep-th/9702161}
\title{New hyper-K\"ahler manifolds by fixing monopoles
\footnote{Phys. Rev. D {\bf 56}, 1120 (1997), hep-th/9702161}} 
\author{Conor J. Houghton\thanks{Electronic address:
    C.J.Houghton@damtp.cam.ac.uk}}
\address{DAMTP, Silver Street,Cambridge, CB3 9EW, United Kingdom}
\date{February 1997} 
\maketitle
\begin{abstract}
  The construction of new hyper-K\"ahler manifolds by taking the
  infinite monopole mass limit of certain
  Bogomol'nyi-Prasad-Sommerfield monopole moduli spaces is considered.
  The one-parameter family of hyper-K\"ahler manifolds due to Dancer
  is shown to be an example of such manifolds. A new family of fixed
  monopole spaces is constructed. They are the moduli spaces of four
  SU$_4$ monopoles, in the infinite mass limit of two of the
  monopoles.  These manifolds are shown to be nonsingular when the
  fixed monopole positions are distinct.
\end{abstract}

\section{Introduction\label{int}}
\ \indent The moduli spaces of Bogomol'nyi-Prasad-Sommerfield (BPS)
monopoles are hyper-K\"ahler manifolds. For charge two SU$_2$ monopoles,
the moduli space is the Atiyah-Hitchin manifold \cite{AH}. For two
distinct monopoles in the maximally broken SU$_3$ theory, the moduli
space is Taub-NUT space \cite{Connell,GL,LWY1}.  Since monopole moduli
spaces have an isometric SO$_3$ action corresponding to rotations of
the monopoles in space, these hyper-K\"ahler manifolds are the only
possible nontrivial four-dimensional monopole moduli spaces
\cite{GP,AH}. In this paper other four-dimensional hyper-K\"ahler manifolds
are derived from monopole moduli spaces by taking the infinite mass
limit of some of the monopole masses, thus fixing the monopole
positions. Fixing monopole positions generally breaks the SO$_3$
isometry.

A one-parameter deformation of the Atiyah-Hitchin manifold is known,
\cite{Dancer2,Dancer3}. It was constructed using the hyper-K\"ahler
quotient \cite{HKLR}. These hyper-K\"ahler manifolds are
reinterpreted as BPS monopole moduli spaces, with one monopole fixed.
A moduli space of BPS monopoles with two fixed monopoles is then
considered. By constructing these moduli spaces via a hyper-K\"ahler
quotient, they are proven to be nonsingular when the fixed monopoles
are fixed at different points in space.

This paper is organized as follows. The Nahm formulation is reviewed in
Sec. II. In Sec. III, Dancer's one-parameter
family of hyper-K\"ahler manifolds is discussed. The moduli spaces with
two fixed monopoles are introduced in Sec. IV and their
nonsingularity is demonstrated in Sec. V. Other fixed
monopole spaces are described in Sec. VI. The paper
concludes in Sec. VII with some remarks about the
applications of the new hyper-K\"ahler manifolds to three-dimensional
supersymmetric theories. There is an Appendix concerning SU$_4$ monopoles.

\section{Nahm data and BPS monopoles}
\ \indent A BPS monopole is a pair $(\Phi,A_i)$ satisfying the
Bogomol'nyi equation. The Higgs field $\Phi$ is an su$_n$ valued scalar
field and $A_i$ is the gauge potential. There is an SU$_n$ gauge
action on these fields, broken by the asymptotic Higgs field. If
SU$_n$ is broken to the maximal torus U$_1^{n-1}$ the Higgs field at
infinity is required to lie in the gauge orbit of 
\be
\Phi_{\infty}=i\mbox{diag}(t_1,t_2,\ldots,t_n).
\ee
By convention $t_1<t_2< \ldots < t_n$ and, since $\Phi$ is traceless,
$t_1+t_2+\ldots+t_n=0$. Because of the asymptotic condition on $\Phi$,
it gives a map from the large sphere at infinity into the quotient
space 
\be\mbox{orbit}_{\mbox{SU}_n}\Phi_{\infty}=\mbox{SU}_n/\mbox{U}_1^{n-1}.\ee
Since $\pi_2(\mbox{SU}_n/\mbox{U}_1^{n-1})={\bf Z}^{n-1}$
the moduli space of monopoles is divided, topologically, into sectors
labelled by $n-1$ integers, $k_i$, called topological charges. The maximal
torus of SU$_n$ is generated by the Cartan space and the matrix
$\Phi_{\infty}$ defines a direction in this Cartan space. This
direction picks out a unique set of simple roots in the Cartan space;
those whose inner product with $\Phi_{\infty}$ is positive. Each U$_1$
in the maximal torus is generated by one of these simple roots. The
$k_i$ are then ordered by the requirement that adjacent $k_i$'s
correspond to non-orthogonal roots. A monopole with topological charge
$(k_1,k_2,\ldots,k_{n-1})$ is called a
\mbox{$(k_1,k_2,\ldots,k_{n-1})$ mono}\-pole. 
                                     
BPS monopoles can be constructed from Nahm data \cite{N}.
In the case of SU$_2$ the space of Nahm data and the moduli space of
monopoles are proven in \cite{HB} to be diffeomorphic and in
\cite{Nak2} this diffeomorphism is proven to be isometric. The spaces
are proven in \cite{HM} to be diffeomorphic when the unbroken symmetry
group is Abelian. It is generally believed that the two spaces are
isometric in all cases and in this paper this is assumed to be
true.

The Nahm data corresponding to a
\mbox{$(k_1,k_2,\ldots,k_{n-1})$ mono}\-pole are a triplet of
skew-hermitian matrix functions defined over the interval $[t_1,t_n]$.
The $t_1< t_2< \ldots < t_{n}$ subdivide the interval into $n-1$
abutting subintervals. For a
\mbox{$(k_1,k_2,\ldots,k_{n-1})$ mono}\-pole a skyline diagram is drawn:
a step function over the interval whose height on the $i$'th
subinterval is $k_i$. For example, a \mbox{$(3,1,2)$ mono}\-pole in an
SU$_4$ theory has diagram 
\be
\begin{array}{c}
\epsfig{file=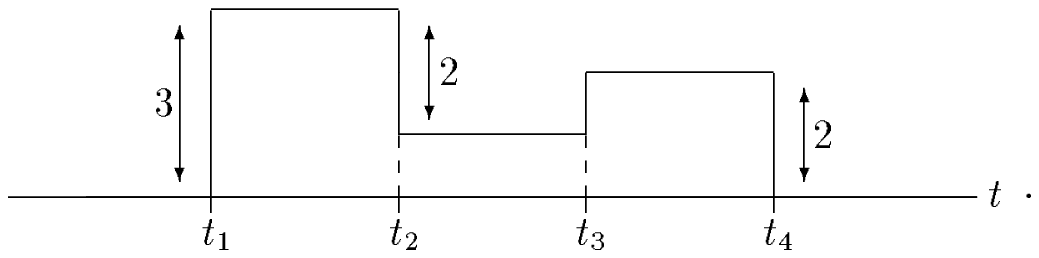,bbllx=155pt,bblly=700pt,bburx=460pt,bbury=780pt,
width=305pt}
\end{array}
\ee

The Nahm triplet is a triplet of square matrix functions of $t$ of
different size over different subintervals. The size of the Nahm
matrices in a subinterval is given by the height of the skyline in that
interval. The matrices must satisfy the Nahm equations in each
subinterval. The Nahm equations are
\be \frac{dT_1}{dt}=[T_2,T_3] \label{Neqn}\ee
and two others by cyclic permutations of 1, 2 and 3. 

There are boundary conditions relating the Nahm matrices in abutting
subintervals. For the purpose of explaining these conditions lets us
consider the skyline diagram
\be
\begin{array}{c}
\epsfig{file=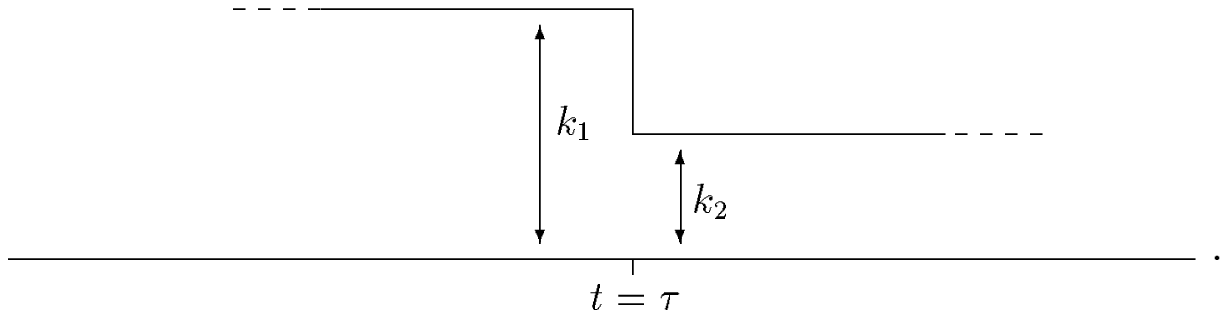,bbllx=135pt,bblly=685pt,bburx=500pt,bbury=790pt,
width=365pt}
\end{array}
\ee
\noindent The skyline is $k_1$ high to the left of $\tau$ and $k_2<k_1$ high to
the right of it. Thus, the Nahm triplet, $(T_1(t),T_2(t),T_3(t))$, is
a triplet of $k_1\times k_1$ matrices over the left interval and of
$k_2\times k_2$ matrices over the right interval. As $t$ approaches
$\tau$ from the left, it is required that 
\be
T_i(s)=\left(\begin{tabular}{cc}$R_i/s+O(1)$ &
      $O(s^{(m-1)/2})$ \\$O(s^{(m-1)/2})$ &
      $T_i^{\prime}+O(s)$
    \end{tabular}\right)\label{frombelow}
\ee
where $s=\tau-t$, $m=k_1-k_2$ and the $k_2\times k_2$ matrix
$T_i^{\prime}$ is the nonsingular limit of the right interval
Nahm data at $t=\tau$.  The $m\times m$ residue matrices $R_i$ in
 (\ref{frombelow}) must form the irreducible $m$-dimensional
representation of su$_2$. Since the one-dimensional representation
is trivial, there is no singularity when $m=1$. When $k_1$ is less
than $k_2$, the conditions are almost the same, again there is a
pole with residue matrices forming the $m=(k_2-k_1)$-dimensional
representation of su$_2$ and the $k_1\times k_1$ data are
submatrices of the $k_2\times k_2$ data at the boundary.  The
situation when $k_1=k_2$ is very different; that case is not
required in this paper.

When some of the $t_i$'s in the asymptotic Higgs field are coincident,
the residual gauge symmetry is enhanced. If two coincide, one U$_1$
factor is replaced by an SU$_2$ factor. If three coincide, two U$_1$'s
are lost and an SU$_3$ gained. Generally the unbroken group is
U$_1^r\times$K where K is a rank $n-r-1$ semisimple Lie group.
Since $\pi_2(\mbox{SU}_n/(\mbox{U}_1^k\times \mbox{K}))={\bf Z}^r$ monopole solutions in
theories with non-Abelian residual symmetries have fewer topological
charges. However, the monopole solutions still have $n-1$ integer
labels. Some of these integers are the usual topological charges. The
rest are what are known as holomorphic charges.

The role of the holomorphic charges is subtle. If two $t_i$'s are
coincident, there is a zero thickness subinterval in the Nahm
interval. The boundary conditions for Nahm data in this situation can
be described in terms of those explained above, by formally imagining
the zero thickness subinterval as the zero thickness limit of a
subinterval of finite thickness. The Nahm data on this subinterval
become irrelevant in the limit, but the height of the skyline on
vanishing subintervals affects the matching condition between the Nahm
matrices over the subintervals on either side.

An example is SU$_3$ broken to U$_1^2$. A
\mbox{$(2,1)$ monopole} has skyline
\be
\begin{array}{c}
\epsfig{file=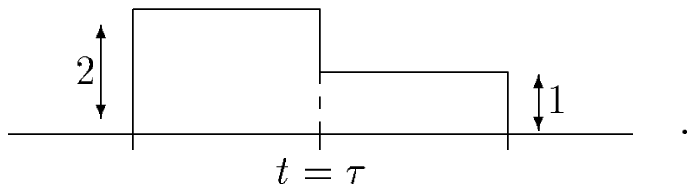,bbllx=205pt,bblly=715pt,bburx=415pt,bbury=775pt,
width=210pt}
\end{array}
\label{21}\ee 
\noindent The Nahm data are $2\times 2$ in the left interval and $1\times 1$ in
the right interval. The Nahm equations (\ref{Neqn}) dictate that
$1\times 1$ data are constant. Therefore, the right interval triplet
is a triplet of imaginary numbers. These numbers are $i$ times the
cartesian coordinates of the ( ,1) part of the
\mbox{$(2,1)$ monopole}. The boundary conditions imply that the
$2\times 2$ data are nonsingular at the boundary, $t=\tau$, between
the two intervals and, further, that their entries
$T_i(\tau)_{_{2,2}}$ are the $1\times 1$ data.  The $2\times 2$ data
are singular on the left boundary of the interval and the residues
there form an irreducible representation of su$_2$. Letting the right
hand interval vanish, a SU$_3$ monopole with topological charge two
and holomorphic charge one is obtained. Holomorphic charges are
distinguished from topological charges by square bracketing them.
Thus, this monopole is a \mbox{(2,[1]) mono}pole and it has skyline
\be 
\begin{array}{c}
\epsfig{file=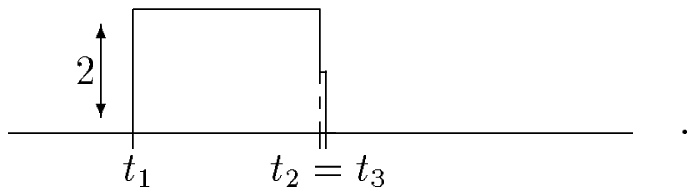,bbllx=205pt,bblly=715pt,bburx=415pt,bbury=775pt,
width=210pt}
\end{array}
\label{2s1s}\ee
The Nahm data are $2\times2$ matrices with a pole on the left boundary but not
on the right one. 

In contrast, a \mbox{$(2,0)$ monopole} has skyline
\be 
\begin{array}{c}
\epsfig{file=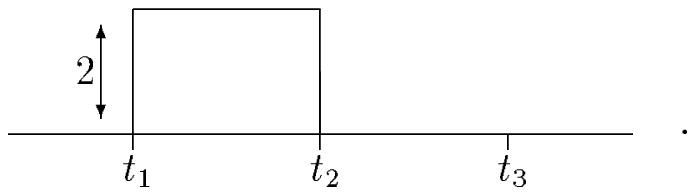,bbllx=205pt,bblly=715pt,bburx=415pt,bbury=775pt,
width=210pt}
\end{array}
\ee
The Nahm data are $2\times 2$ matrices over the left hand
subinterval and have poles at $t_1$ and $t_2$. There is no data over
the right hand subinterval. These data are identical to SU$_2$
2-monopole data and correspond to the embedding of an SU$_2$
2-monopole in SU$_3$. The length of the right hand subinterval does not
affect the Nahm data, there is a pole at both $t_1$ and $t_2$
irrespective of whether $t_2=t_3$ or not. If $t_2=t_3$ the Nahm data
correspond to a \mbox{(2,[0]) monopole}. 

These examples demonstrate how the holomorphic charges determine the
boundary conditions and how these boundary conditions can be derived
by imagining the non-Abelian case as the zero interval thickness limit
of the Abelian case.  It should be noted that different holomorphic
charges do not necessarily correspond to different monopoles or to
different Nahm data. For example, \mbox{(3,[1]) monopoles} can
equally well be called \mbox{(3,[2]) monopoles}. This ambiguity is
discussed, for example, by Weinberg in \cite{W}.

\section{Dancer's family of hyper-K\"ahler manifolds}
\ \indent In \cite{Dancer2}, the moduli space of centered
\mbox{$(2,[1])$ mono}poles is constructed. These monopoles are charge
$(2,[1])$ SU$_3$ monopoles. They have the skyline diagram
(\ref{2s1s}). They are called centered because their Nahm data are
traceless. The moduli space, $M_0^8$, is eight dimensional.  The Nahm
data for such monopoles are a triplet of $2\times2$ traceless
skew-hermitian matrix function over the interval $[-2,1]$. There is a
simple pole at $t=-2$ and the residues there form the irreducible two-dimensional representation of su$_2$. The space of such Nahm triplets,
$M_0^5$, is five dimensional. The whole of $M_0^8$ is generated by the
action of SU$_2$ on these Nahm data.

In the last section, for simplicity, the Nahm data described are gauge
fixed. While the gauge fixed Nahm data are a triplet of matrix
functions, to form the required SU$_2$ orbit of $M_0^5$ the
quadruplet of matrix functions $(T_0,T_1,T_2,T_3)$ is introduced. This
quadruplet is required to satisfy the Nahm equations \be
\frac{dT_1}{dt}+[T_0,T_1]=[T_2,T_3] \label{Neqn2}\ee and two others by
cyclic permutations of 1, 2 and 3.

The introduction of $T_0$ allows a group action to be defined on the
space of $(2,[1])$ Nahm data. If \be _0{\cal G}=\{g\in
C^w([-2,1],U_2):g(-2)=\identity\}\ee and its subgroup \be _0{\cal
  G}_0=\{g\in C^w([-2,1],\mbox{U}_2):g(-2)=g(1)=\identity\}\ee an action of
$g\in$ $_0{\cal G}$ on $(T_0,T_1,T_2,T_3)$ is defined by
\bea &&T_0 \mapsto  gT_0g^{-1}-\frac{dg}{dt}g^{-1}, \label{gact}\\
&&T_i \mapsto gT_ig^{-1},\qquad\qquad(i=1,2,3). \nonumber\eea If
$g\in$ $_0{\cal G}_0$ then the action is a gauge action.  The moduli
space of uncentered Nahm data, $M^{12}$, is the space of gauge
inequivalent data. Furthermore, U$_2=_0\!{\cal G}/_0{\cal G}_0$ and, so, a
U$_2$ action on the data is given by Eq. (\ref{gact}). A hyper-K\"ahler
quotient by the center of this U$_2$ on $M^{12}$ centers the Nahm
data, giving $M_0^8$. The remaining SU$_2$ action can be fixed by
setting $T_0$ to zero, reducing Eq. (\ref{Neqn2}) to Eq. (\ref{Neqn}) and
$M_0^8$ to $M_0^5$.

There is also an SO$_3$ action. 
It both rotates the Nahm triplet as a vector and gauge transforms the
four Nahm matrices. This action is not triholomorphic; it rotates the
complex structures.

The SU$_2$ action on $M_0^8$ is triholomorphic and isometric. This
means that there is an induced moment map, $\mu$, from $M_0^8$ to ${\bf R}^3$
formed by the action of a U$_1$ subgroup of SU$_2$. Dancer's family of
hyper-K\"ahler manifolds is the family of hyper-K\"ahler four manifolds
\be
M(\lambdabf)=\mu^{-1}(\lambdabf)/\mbox{U}_1,\nonumber
\ee
where $\lambdabf\in{\bf R}^3$.
The SO$_3$ action on $M^8_0$ is not an
isometry of $M(\lambdabf)$, rather, it acts on $\lambdabf$ to give an
isometry between $M(\lambdabf)$ and $M(R\lambdabf)$ where 
$R$ is an SO$_3$ matrix. $M({\bm 0})$ is a double cover of the
Atiyah-Hitchin manifold.

The hyper-K\"ahler manifolds $M(\lambdabf)$ are hyper-K\"ahler quotients
of a monopole moduli space.  It is now  shown that they are the
infinite mass limit of another monopole space.  The moment map $\mu$
is known explicitly. If the U$_1$ subgroup is the subgroup which fixes
$i\sigma_3$ when SU$_2$ acts on su$_2$ in the adjoint representation,
the moment map $\mu:M_0^8\rightarrow{\bf R}^3$ given by this U$_1$ action
is \be
\mu:(T_0,T_1,T_2,T_3)\mapsto(-\mbox{trace}(T_1(1)i\sigma_3),-\mbox{trace}(T_2(1)i\sigma_3),-\mbox{trace}(T_3(1)i\sigma_3)).
\ee The level set $\mu^{-1}(\lambdabf)$ consists of Nahm data whose
entries $T_i(1)_{_{2,2}}$ are $i\lambda_i/2$ at $t=1$.  For
\mbox{$(2,1)$ mono}poles, (\ref{21}), the data in the right-hand
interval are given by the $T_i(1)_{_{2,2}}$ entries of the left-hand
Nahm data at the boundary. Thus, the hyper-K\"ahler manifolds
$M(\lambdabf)$ are the moduli spaces of \mbox{$(2,1)$ monopoles} with
the \mbox{$(\;,1)$ monopole} fixed. The \mbox{$(\;,1)$ mono}\-pole can
be fixed by taking its infinite mass limit. The monopole mass is
proportional to the length of the corresponding interval, so this
limit is
\be 
\begin{array}{c}
\epsfig{file=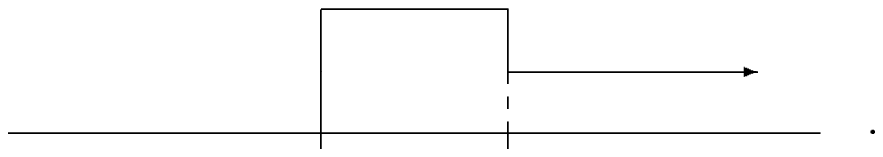,bbllx=205pt,bblly=715pt,bburx=415pt,bbury=775pt,
width=210pt}
\end{array}
\label{2r1r}\ee
The vector $\lambdabf$ is now related to the
position of the \mbox{$(\;,1)$ monopole}: the monopole whose position
is fixed. When the position of the \mbox{$(\;,1)$ monopole} is fixed
in the center, the relative metric of the \mbox{$(2,\;)$ monopole} is
Atiyah-Hitchin. That is not surprising. It has been noted,
\cite{Bi2,HSd}, that if three SU$_2$ monopoles are lined up, with
suitable relative phases, the metric is Atiyah-Hitchin.

An advantage of this description of $M(\lambdabf)$ is that its
asymptotic behaviour may be calculated using the methods of
\cite{M2,GM2,LWY2}, that is by approximating the monopoles by point
particles and calculating their long range interactions.
This yields a
purely kinetic Lagrangian for the motion of the well separated
monopoles and, hence, an asymptotic metric. This
metric is  
\be ds^2=g_{ij}d{\bm x}_i\cdot d{\bm x}_j+g_{ij}^{-1}(d\chi_i+{\bm
W}_{ik}\cdot d{\bm x}_k)(d\chi_j+{\bm W}_{jl}\cdot d{\bm x}_l)\ee 
where, with no sum on repeated indices,
\begin{eqnarray}
g_{jj}&=&m_{j}-\sum_{i\not=j}\frac{\alpha_{ij}}{r_{ij}}\label{asymgen}\\
g_{ij}&=&\frac{\alpha_{ij}}{r_{ij}},\qquad(i\not=j)\nonumber\\
{\bm W}_{jj}&=&-\sum_{i\not=j}{\bm w}_{ij},\nonumber\\
{\bm W}_{ij}&=&{\bm w}_{ij},\qquad(i\not=j)\nonumber
\end{eqnarray}
and ${\bm x}_i$, $\chi_i$ and $m_j$ are the spacial coordinates, phases
and masses of
the monopoles; these are all well defined in the point particle
approximation. A Dirac potential ${\bm w}({\bm r})$ satisfies
\be \nabla_{{\bm r}}\times{\bm w}=-\frac{{\bm r}}{r^3}.\ee
In  Eq. (\ref{asymgen}) $r_{ij}=|{\bm x}_i-{\bm x}_j|$ and ${\bm w}_{ij}$ is
the corresponding Dirac potential. If the $i$ and $j$ monopoles have
the same U$_1$ charge then $\alpha_{ij}=1$ and if they correspond to
adjacent U$_1$'s, $\alpha_{ij}=-1/2$, otherwise it is zero. 

For \mbox{$(2,1)$ monopoles}  
\be
\begin{array}{c}
\epsfig{file=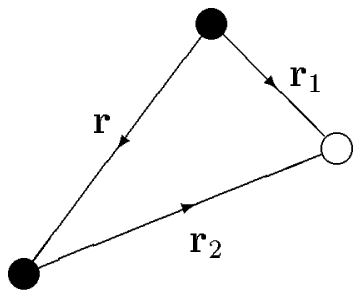,bbllx=245pt,bblly=690pt,bburx=355pt,bbury=785pt,
width=110pt}
\end{array}
\ee the mass of the two \mbox{$(2,\;)$ mono}poles is chosen to be one and that
of the \mbox{$(\;,1)$ monopole} to be $m$. Then 
\be
g_{ij}=\left( \begin{array}{ccc}
    1-\frac{1}{r}+\frac{1}{2r_1}&\frac{1}{r}&-\frac{1}{2r_1}\\
    \frac{1}{r}&1-\frac{1}{r}+\frac{1}{2r_2}&-\frac{1}{2r_2}\\

    -\frac{1}{2r_1}&-\frac{1}{2r_2}&m+\frac{1}{2r_1}+\frac{1}{2r_2}\end{array}
\right).\ee 
In the $({\bm r},\frac{1}{2}({\bm x}_1+{\bm x}_2),{\bm
  x}_3-\frac{1}{2}({\bm x}_1+{\bm x}_2))$ basis this becomes 
\be g^{\prime}_{ij}=\left(\begin{array}{ccc}
    \frac{1}{2}-\frac{1}{r}+\frac{1}{8r_1}+\frac{1}{8r_2}&0&\frac{1}{4r_2}-\frac{1}{4r_1}\\
    0&m+2&m\\
    \frac{1}{4r_2}-\frac{1}{4r_1}&m&m+\frac{1}{2r_1}+\frac{1}{2r_2}
\end{array}
\right).
\ee
Thus, taking the infinite mass limit, the asymptotic metric on
$M(\lambdabf)$ is \bea
ds^2&=&V_1d{\bm r}\cdot d{\bm r}+V_2^{-1}(d\chi+{\bm W}\cdot d{\bm r})^2\label{Dgiasym}\\
V_1&=&\frac{1}{2}-\frac{1}{r}+\frac{1}{8r_1}+\frac{1}{8r_2}\nonumber\\
V_2&=&1-\frac{1}{r}+\frac{1-4r_{1}r_{2}}{8r_{1}r_{2}+2r_{1}+2r_{2}}\nonumber\\
{\bm W}&=&-{\bm w}+\frac{1}{8}{\bm w}_1+\frac{1}{8}{\bm
  w}_2.\nonumber\eea 
This metric is singular as $r\rightarrow 0$. It is only valid for large $r$. 

The asymptotic metric for $M({\bm 0})$ is found by placing the fixed
monopole at the center of mass of the two unfixed monopoles and, thus,
by substituting $r_{1}=r_{2}=r/2$ and ${\bm w}_1={\bm w}_2=2{\bm w}$
in Eq. (\ref{Dgiasym}). Making these substitutions reduces Eq. (\ref{Dgiasym})
to the asymptotic form of the Atiyah-Hitchin metric.

\section{A new family of hyper-K\"ahler manifolds}
\ \indent Another advantage of this description is that it immediately
suggests a new family of four-dimensional hyper-K\"ahler manifolds,
$N(\lambdabf,\mubf)$. In Sec. III, it is shown that
$M(\lambdabf)$ is a fixed monopole space derived from the moduli space
of charge $(2,1)$ SU$_3$ monopoles. This suggests that a new family of
hyper-K\"ahler manifolds could be constructed by fixing monopoles in
the moduli space of charge $(1,2,1)$ SU$_4$ monopoles. A
\mbox{$(1,2,1)$ mono}pole has skyline 
\be 
\begin{array}{c}
\epsfig{file=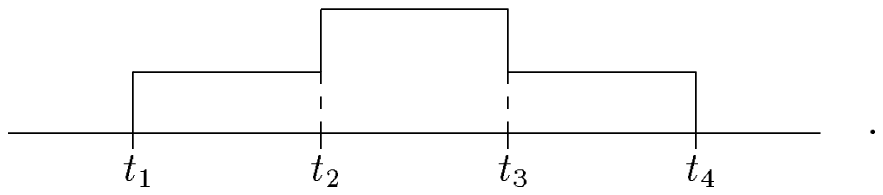,bbllx=150pt,bblly=710pt,bburx=410pt,bbury=780pt,
width=260pt}
\end{array}
\label{121}\ee
The corresponding Nahm data are $2\times 2$ matrices in the middle
subinterval and $1\times 1$ matrices in the left and right
subintervals. The Nahm data in the left subinterval are equal to the
entries $T_i(t_2)_{_{2,2}}$ of the $2\times 2$ data, the Nahm data in
the right subinterval are equal to the entries $T_i(t_3)_{_{2,2}}$. All
the Nahm data are analytic.

The limit where the subintervals $[t_1,t_2]$ and
$[t_3,t_4]$ become infinitely long gives the $(1,2,1)$
fixed monopole spaces:
\be 
\begin{array}{c}
\epsfig{file=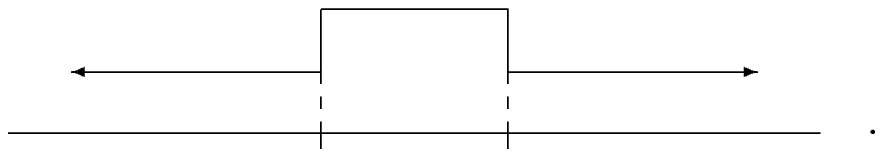,bbllx=150pt,bblly=710pt,bburx=410pt,bbury=780pt,
width=260pt}
\end{array}
\label{r1r2r1r}\ee
They are labelled by two vectors, $\lambdabf$ and $\mubf$, the
positions of the two fixed monopoles: the \mbox{$(1,\;,\;)$ monopole}
and the \mbox{$(\;,\;,1)$ monopole}. These spaces are denoted
$N(\lambdabf,\mubf)$.  The SO$_3$ action on the charge $(1,2,1)$
moduli space is isometric and rotates the two vectors $\lambdabf$ and
$\mubf$. In the infinite mass limit of the \mbox{$(1,\;,\;)$ monopole}
and the \mbox{$(\;,\;,1)$ monopole}, the action of some $R\in$SO$_3$
gives an isomorphism between $N(\lambdabf,\mubf)$ and
$N(R\lambdabf,R\mubf)$. Thus, $N(\lambdabf,\mubf)$ is a
three-parameter family of hyper-K\"ahler manifolds. If $\lambdabf$ and
$\mubf$ are parallel then a U$_1$ subgroup of the SO$_3$ action fixes
$N(\lambdabf,\mubf)$ and so $N(\lambdabf,\mubf)$ has a U$_1$ isometry.

Using the same methods as in Sec. III, the asymptotic form of the
$N(\lambdabf,\mubf)$ metric can be calculated. It is 
\bea
ds^2&=&V_1d{\bm r}\cdot d{\bm r}+V_2^{-1}(d\chi+{\bm W}\cdot d{\bm r})^2\label{newgiasym}\\
V_1&=&\frac{1}{2}-\frac{1}{r}+\frac{1}{8r_{11}}+\frac{1}{8r_{12}}
+\frac{1}{8r_{21}}+\frac{1}{8r_{22}}\nonumber\\
V_2&=&1-\frac{1}{r}+\frac{1}{2}\frac{r_{11}r_{21}+
  r_{11}r_{22}+r_{12}r_{21}+r_{12}r_{22}-4r_{11}r_{12}r_{21}r_{22}}{4r_{11}r_{12}r_{21}r_{22}+r_{11}r_{12}r_{21}+r_{11}r_{12}r_{22}+r_{11}r_{21}r_{22}+r_{12}r_{21}r_{22}}\nonumber\\
{\bm W}&=&-{\bm w}+\frac{1}{8}{\bm w}_{11}+\frac{1}{8}{\bm
  w}_{12}+\frac{1}{8}{\bm w}_{21}+\frac{1}{8}{\bm w}_{22}\nonumber\eea
where everything is defined as before, except that now there are two
fixed monopoles and the distances from the two
\mbox{$(\;,2,\;)$ monopoles} to the first of these have been denoted
by $r_{11}$ and $r_{21}$ and the distances to the second by $r_{12}$ and
$r_{22}$.
\be 
\begin{array}{c}
\epsfig{file=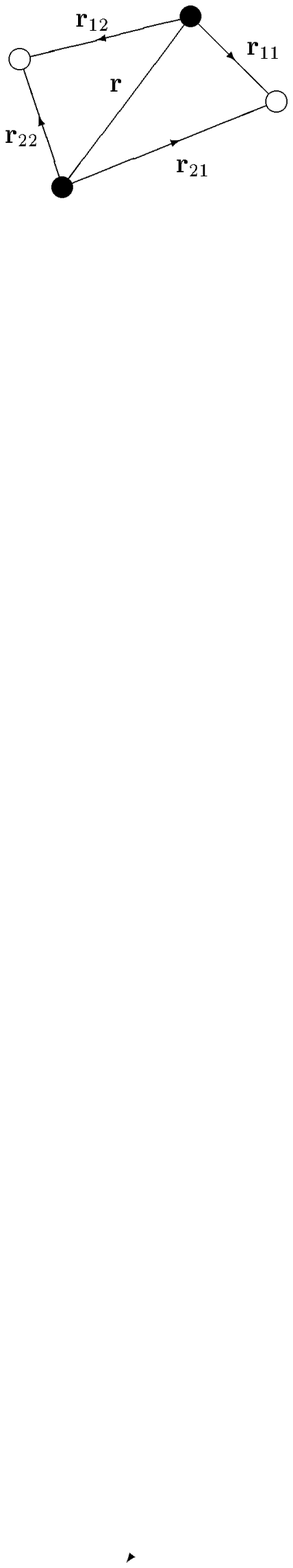,bbllx=225pt,bblly=690pt,bburx=370pt,bbury=785pt,
width=145pt}
\end{array}
\ee
Examining the asymptotic formula, it is interesting to see how flat 
the $N(\lambdabf,\mubf)$ metrics are. All the metrics are flat up to
the second order in $1/r$. 

\section{nonsingularity of the new hyper-K\"ahler manifolds}
\ \indent It is not clear from the discussion in Sec. IV
that the \mbox{$(1,2,1)$ moduli} space remains nonsingular as the
masses of the \mbox{$(1,\;,\;)$ monopole} and the
\mbox{$(\;,\;,1)$ monopole} become infinite. Dancer's family;
$M(\lambdabf)$, is known to be nonsingular because it can be
constructed using a hyper-K\"ahler quotient. In imitation of this,
$N(\lambdabf,\mubf)$ is constructed in this section by
hyper-K\"ahler quotient of the moduli space of
\mbox{$([1],2,[1])$ mono}poles. These monopoles are SU$_4$ monopoles
of topological charge two in the theory where the residual symmetry is
SU$_2\times$U$_1\times$SU$_2$. The skyline diagram is \be
\begin{array}{c}
\epsfig{file=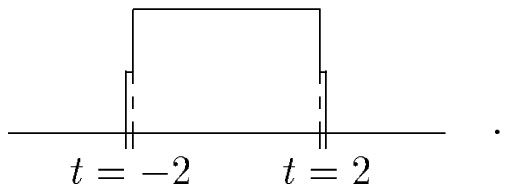,bbllx=205pt,bblly=715pt,bburx=415pt,bbury=775pt,
width=210pt}
\end{array}
\label{s1s2s1s}\ee
The Nahm data are $2\times 2$ matrices analytic over the whole
interval $[2,2]$. There are two commuting SU$_2$ actions, one at each
boundary. These data correspond to \mbox{$([1],2,[1])$ mono}\-poles.
In this section, the moduli space of \mbox{$([1],2,[1])$ mono}\-poles
is used to construct $N(\lambdabf,\mubf)$ in the same way as the
moduli space of \mbox{$(2,[1])$ monopoles} is used to construct
$M(\lambdabf)$. It is found that the manifold $N(\lambdabf,\mubf)$ is
free of singularities as long as $\lambdabf\not=\mubf$.

The charge ([1],2,[1]) Nahm data are quadruplets $(T_0,T_1,T_2,T_3)$
satisfying the Nahm equations (\ref{Neqn2}) and acted on by the gauge
group
\be _0{\cal G}_0=\{g\in C^w([-2,2],\mbox{U}_2):g(-2)=g(2)=\identity\}.\ee
The two larger groups,
\bea &&_0{\cal G}=\{g\in C^w([-2,2],\mbox{U}_2):g(-2)=\identity\},\\
  &&{\cal G}_0=\{g\in C^w([-2,2],\mbox{U}_2):g(2)=\identity\}\eea
are defined. These are subgroups of ${\cal G}=\{g\in C^w([-2,2],\mbox{U}_2)\}$.

Two U$_2$ actions are given by $_0{\cal G}/_0{\cal G}_0$ and ${\cal
  G}_0/_0{\cal G}_0$. These actions commute. The whole U$_2\times $U$_2$
action is the ${\cal G}/_0{\cal G}_0$ action. The
center is U$_1\times$U$_1$. The Nahm data are fixed under the central
element represented by the constant function $g(t)=e^{i\theta
  }\identity_2$.  The element represented by $g(t)=e^{i\theta
  t}\identity_2$ sends $(T_0,T_1,T_2,T_3)$ to
$(T_0-i\theta\identity_2,T_1,T_2,T_3)$ and generates the vector field
$(-i\identity_2,0,0,0)$. The hyper-K\"ahler quotient by this action
centers the Nahm data. This space of centered data is called
$N_0^{12}$. It is twelve dimensional. It has an isometric
triholomorphic SU$_2\times $SU$_2$ action. There is also an SO$_3$
action, which rotates $(T_1,T_2,T_3)$ as a three vector and commutes
with the SU$_2\times $SU$_2$ action.

A U$_1\times $U$_1$ subgroup of the SU$_2\times$SU$_2$ is represented by
the elements \be \alpha(t)=e^{\frac{i\theta}{4} (t+2) \sigma_3},\qquad
\beta(t)=e^{\frac{i\theta}{4} (2-t) \sigma_3}.\label{alphabeta} \ee
The moment map, $\mu:N_0^{12}\rightarrow{\bf R}^3\times{\bf R}^3$, for the action
of this subgroup is \be
\mu:(T_0,T_1,T_2,T_3)\mapsto(\lambdabf,\mubf)\ee where \be
\lambdabf=(-\mbox{trace}(T_1(-2)i\sigma_3),
-\mbox{trace}(T_2(-2)i\sigma_3),-\mbox{trace}(T_3(-2)i\sigma_3))\ee
and \be
\mubf=(-\mbox{trace}(T_1(2)i\sigma_3),-\mbox{trace}(T_2(2)i\sigma_3),-\mbox{trace}(T_3(2)i\sigma_3)).\ee
By the same argument as in Sec. III, $N_0^{12}$ reduces
to $N(\lambdabf,\mubf)$ under the hyper-K\"ahler quotient: \be
N(\lambdabf,\mubf)=\mu^{-1}(\lambdabf,\mubf)/\mbox{U}_1\times
\mbox{U}_1\label{hkqreal} \ee

The condition that $\lambdabf$ and $\mubf$ must satisfy, in order for
the U$_1\times $U$_1$ action to be free, are now needed.
These are the 
conditions for the nonsingularity of the $N(\lambdabf,\mubf)$. 

To apply these conditions it is necessary to solve the Nahm
equations. Using the ${\cal G}$ action, $T_0$ is gauged to zero. This
leaves an eight-dimensional space acted on by constant elements of
${\cal G}$ and by the SO$_3$ action. By acting with the SO$_3$ the $t$
invariants: $\mbox{trace}(T_1T_2)$, $\mbox{trace}(T_2T_3)$ and
$\mbox{trace}(T_3T_1)$, can be set to zero.  This means that if the
$T_i$ are written as
\be T_i=\half i f_i {\bm n}_i\cdot
\sigmabf,
\label{SO3fix}
\ee
where $i=1,2,3$ and the ${\bm n}_i$ are constant orthonormal vectors and so the functions 
$f_1$, $f_2$ and $f_3$ satisfy 
\be \frac{df_1}{dt}=f_2f_3\label{Eteq}\ee
and two others given by cyclic permutations of 1, 2 and 3. 
The SO$_3$ action are completely fixed by requiring that

\be f_1^2\le f_2^2 \le f_3^2. \label{order}\ee

The remaining group action is that of constant elements of ${\cal G}$.
It is fixed by setting ${\bm n}_1=(1,0,0)$, ${\bm n}_2=(0,1,0)$ and
${\bm n}_3=(0,0,1)$.  The resulting subspace of the moduli space
$N_0^{12}$ is called $N^3$.  Since the SO$_3$ action on
$N_0^{12}$ is not free, $N^3$ is not a manifold.  Equations
(\ref{Eteq}) are the well known Euler top equations and are solved
in terms of Jacobi elliptic functions as 
\bea
&&f_1(t)=\pm\frac{Dcn_k(D(t+\tau))}{sn_k(D(t+\tau))},\label{Et1}\\
&&f_2(t)=\pm\frac{Ddn_k(D(t+\tau))}{sn_k(D(t+\tau))},\nonumber\\
&&f_3(t)=\pm\frac{D}{sn_k(D(t+\tau))},\nonumber
\eea
where $0\le k\le 1$ is the elliptic modulus, $D$ and $\tau$ are
arbitrary real constants and the signs are all minus or exactly two of
them are plus.  Analyticity of the data requires that $\tau>2$ and
$D(\tau+2)<2K(k)$, where $4K(k)$ is the period of $sn_k$.  Further
solutions are found by changing the sign of all three $f_i$'s and
sending $t$ to $-t$. The analyticity requirements on these further
solutions are that $\tau<2$ and $D(\tau+2)<2K(k)$.  This exhausts all
the solutions consistent with the various conditions which have been
imposed.

The action of $\alpha(t)$ and $\beta(t)$ is given
by Eq. (\ref{alphabeta}). Since $T_0$ is zero on $N^3$ the only element in
the group generated by $\alpha(t)$ and $\beta(t)$ which could have a
fixed point in $N^3$ is the constant one
$\alpha\beta(t)=e^{i\theta\sigma_3}$. For $\alpha\beta$ to have a
fixed point in $N^3$ it is necessary and sufficient that
$f_1(0)=f_2(0)=0$. This only occurs if $k=1$ and $\tau=\infty$. The
solutions Eq. (\ref{Et1}) are then $f_1(t)=0$, $f_2(t)=0$ and $f_3(t)=D$
and the hyper-K\"ahler quotient gives the space $N((0,0,D),(0,0,D))$.
This means $N(\lambdabf,\mubf)$ with $\lambdabf=\mubf=(0,0,D)$ may
have a singularity. By considering the action on $N^3$ of SU$_2\times
$SU$_2$, it is seen that the only points in $N_0^{12}$ where the action
of $\alpha(t)$ or $\beta(t)$ is not free are those points in the
SO$_3\times(\mbox{constant elements of }{\cal G})$ orbit of the fixed
points occuring in $N^3$. Therefore, the only potentially singular
$N(\lambdabf,\mubf)$ manifolds are $N(\lambdabf,\lambdabf)$.  In the
fixed monopole description, these are the manifolds of coincident
fixed monopoles.

The manifold $N({\bm 0},{\bm 0})$ is singular. This is in contrast
with $M({\bm 0})$ which is a double cover of the Atiyah-Hitchin
manifold. The $N(\lambdabf,\mubf)$ spaces are not deformations of a smooth
SO$_3$ isometric hyper-K\"ahler manifold. It would be interesting to
understand more of the geometry and topology of these spaces.

\section{Other fixed monopole spaces}
\ \indent Following the example of $M(\lambdabf)$ and
$N(\lambdabf,\mubf)$ it is natural to ask whether further nonsingular
fixed monopole spaces might be constructed by fixing larger numbers of
monopoles. For example, a large class of four-dimensional
hyper-K\"ahler manifolds might be derived from the
\mbox{$(k_1,2,k_2)$ mono}\-pole moduli spaces. One might conjecture
that, as long as the \mbox{$(k_1,\;,\;)$ mono}\-poles and the
\mbox{$(\;,\;,k_2)$ mono}\-poles are not fixed in coincident
positions, new multi-parameter families of four-dimensional
hyper-K\"ahler manifolds could result.

More complicated mixtures of fixed and unfixed monopoles could be used
to give fixed monopole spaces of dimensions higher than four. Fixed
charges are distinguished from other charges by enclosing them in
curly brackets. It could be conjectured that for $r>1$ the
$(\{k_1\},l_1,l_2,\ldots,l_r,\{k_2\})$ spaces are nonsingular when
the \mbox{$(k_1,\;,\ldots,\;,\;)$ mono}\-poles and the
\mbox{$(\;,\;,\ldots,\;,k_2)$ mono}\-poles are each fixed so they are
not coincident with monopoles of the same type. 
 
The asymptotic metrics can always be constructed for fixed monopole
spaces using the point monopole methods of \cite{M2,GM2,LWY2}.
Generally, these asymptotic fixed monopole metrics are singular.
This is not the case for the $(\{k\},1)$ space. In the limit of
infinite \mbox{$(k,\;)$ monopole} mass the \mbox{$(k,1)$ mono}\-pole
asymptotic metric is the $k$ center multi-Taub-NUT metric of Hawking
\cite{Hawking}.  The positions of the $k$ centers are the $k$ fixed
monopole positions.  Since the multi-Taub-NUT metric is generically
nonsingular and is the same asymptotically as the $({k},1)$ metric,
it seems likely that they are the same everywhere. Certainly, the
\mbox{$(1,1)$ monopole} metric is known explicitly
\cite{Connell,GL,LWY1} and the $(\{1\},1)$ metric is Taub-NUT. The
\mbox{$(1,1,1)$} metric is also known \cite{LWY2,MKM} and the infinite mass limit
$(\{1\},1,\{1\})$ is two center multi-Taub-NUT. 

Mixtures of fixed,
topological and holomorphic charges might also be considered. An
example is the eight-dimensional space $(\{1\},2,[1])$:
\be 
\begin{array}{c}
\epsfig{file=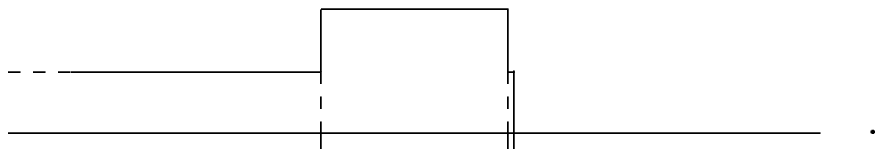,bbllx=150pt,bblly=710pt,bburx=410pt,bbury=780pt,
width=260pt}
\end{array}
\label{r1r2s1s}\ee
It is an interesting space, it has a tri-holomophic SU$_2$ isometry
and an isometric U$_1$ action which rotates the complex structures.

\section{Applications}
\ \indent

The $N(\lambdabf,\mubf)$ are gravitational instantons. Gravitational
instantons are asymptotically flat solutions of the vacuum Einstein
equations. All asymptotically flat four-dimensional hyper-K\"ahler
manifolds are gravitational instantons. As noted earlier,
$N(\lambdabf,\mubf)$ approaches flat space very rapidly.

Fixed monopole spaces are relevant to (2+1)-dimensional quantum field
theories.  In a celebrated recent paper, \cite{HW}, Hanany and Witten
propose a correspondence between three-dimensional supersymmetric
gauge theories and moduli spaces of magnetic monopoles. In the
language of \cite{HW} the fixed monopole spaces correspond to brane
configurations in which some of the threebranes are infinitely
extended in the direction along which the fivebranes are separated.  Thus,
$N(\lambdabf,\mubf)$ corresponds to the configuration 
\be 
\begin{array}{c}
  \epsfig{file=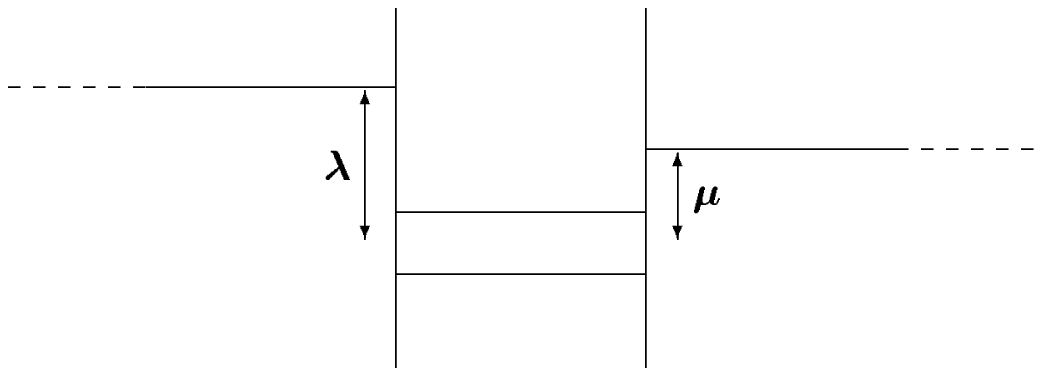,bbllx=150pt,bblly=630pt,bburx=450pt,bbury=750pt,
width=300pt}
\end{array}
\ee
and to quantum field theories with hypermultiplets of masses
$\lambdabf$ and $\mubf$. The Dancer space $M(\lambdabf)$ corresponds
to 
\be 
\begin{array}{c}
  \epsfig{file=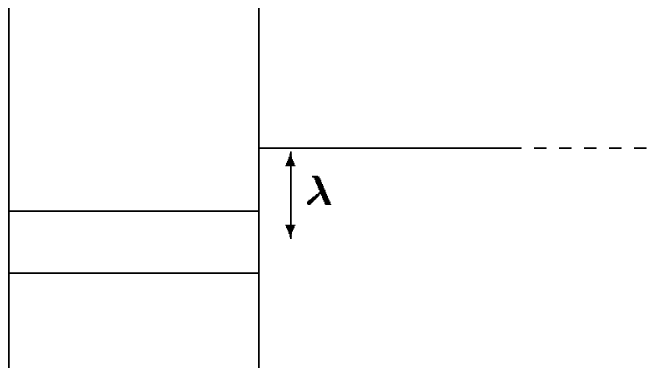,bbllx=150pt,bblly=630pt,bburx=450pt,bbury=750pt,
width=300pt}
\end{array}
\ee
and to quantum field theories with hypermultiplets of mass
$\lambdabf$. These correspondences are described generally in \cite{HW}. The
reinterpretation of $M(\lambdabf)$ as a fixed monopole moduli space gives an
explanation, in the spirit of \cite{HW}, of the appearance of
$M(\lambdabf)$ in \cite{SW}.

\section*{Acknowledgments}

I would like to thank Nick Manton and Paulina Rychenkova for useful
discussions and for helpful comments on earlier versions of this paper. I
thank the EPSRC and the British Council for
financial assistance.

\section*{Appendix: a note on \mbox{$([1],2,[1])$ mono}\-poles}
\ \indent The moduli space of \mbox{$([1],2,[1])$ mono}\-poles was
used in Sec. V to prove the nonsingularity of
$N(\lambdabf,\mubf)$. The discussion in Sec. V would also
be useful in studying \mbox{$([1],2,[1])$ mono}\-poles {\sl per se}.
All \mbox{$([1],2,[1])$ mono}\-poles are $D_2$ symmetric about some
axes. The monopole can be orientated by imposing $D_2$ symmetry about
particular axes. By imposing $D_2$ symmetry about the Cartesian axes,
the monopoles are restricted to a three-dimensional geodesic
submanifold of the moduli space, this is called \X. The space
$N^3$ of Nahm data described above is the quotient of the full moduli
space by the full SO$_3$ action and since this action is not free,
$N^3$ is not a manifold. Instead of quotienting the space of Nahm data
by SO$_3$, $D_2$ symmetry is imposed on it, giving $\X$. The $D_2$
symmetry conditions are identical to Eq. (\ref{SO3fix}) but without the
ordering condition (\ref{order}).  Thus, $\X$ is composed of the six
copies of $N^3$ obtained by permuting the inequality (\ref{order}).
These copies are joined at the planes where two of the $f_i$'s are
equal. These data, where two of the $f_i$'s are equal, correspond to
axially symmetric monopoles. The planes intersect on the lines of
spherical symmetry. An example of a line of spherical symmetry is \be
f_1(t)=f_2(t)=f_3(t)=-\frac{1}{t+\tau} \ee where $\tau>2$.

There are  exceptional lines in $\X$ given by letting $k=1$
and taking $\tau$ to infinity. These lines are notable in the context
of Sec. V as the fixed points of the U$_1\times$U$_1$
action.  These are the lines where one $f_i$ is
constant and the other two are zero. They meet at the point where all
three $f_i$ are zero. These lines correspond to the exceptional
\mbox{$([1],2,[1])$ mono}\-poles produced by embedding two SU$_2$
\mbox{1-mono}\-poles. 

In their paper \cite{DL}, Dancer and Leese studied the head on
collision of \mbox{$(2,[1])$ mono}\-poles. These collisions are
described by geodesics on a two-dimensional manifold that they call
$Y$. $\X$ is the analog of $Y$ for
\mbox{$([1],2,[1])$ mono}\-poles. The boundaries of $\X$ occur when
$(D,\tau)$ attain the bounds imposed by analyticity. When $(D,\tau)$
attain these bounds, the Nahm data has a pole at one or the other end.
This means these boundaries are actually copies of the space $Y$. In
fact, the whole of $\X$ has eight copies of $Y$ at its boundaries.

We can picture $\X$. Take the ${\bf R}^3$ Cartesian axes and thicken them.
Divide the surfaces of these thickened axes by tracing their
intersections with the $xy$, $yz$ and $zx$ planes. The eight surface
elements bounded by these lines are the eight copies of $Y$. The
interior of the thick\-ened axes is \X. The Cartesian axes themselves
are the lines of embedded monopoles. The origin is the spherical
embedded monopole. The intersections of the six planes $x=\pm y$, $y=
\pm z$ and $z= \pm x$ with $\X$ are the planes of axially symmetric
monopoles.  The lines $x=\pm y=\pm z$ are the lines of spherically
symmetric monopoles. This picture of $\X$ is not metrically correct.

\end{document}